\documentclass[preprint,english,superscriptaddress,titlepage]{revtex4}%
\usepackage{amsfonts}
\usepackage[T1]{fontenc}
\usepackage[latin9]{inputenc}
\usepackage{amsmath}
\usepackage{amssymb}
\usepackage{babel}
\usepackage{graphicx}%

\begin{document}
\title{Relativistic time-dependent quantum dynamics across supercritical barriers for
Klein-Gordon and Dirac particles}
\author{M.\ Alkhateeb}
\affiliation{Laboratoire de Physique Th\'eorique et Mod\'elisation, CNRS Unit\'e 8089, CY
Cergy Paris Universit\'e, 95302 Cergy-Pontoise cedex, France}
\author{X. Gutierrez de la Cal}
\affiliation{Departmento de Qu\'imica-F\'isica, Universidad del Pa\' is Vasco, UPV/EHU,
Leioa, Spain}
\author{M. Pons}
\affiliation{Departamento de F\'isica Aplicada I, Universidad del Pa\' is Vasco, Spain}
\author{D. Sokolovski}
\affiliation{Departmento de Qu\'imica-F\'isica, Universidad del Pa\' is Vasco, UPV/EHU,
Leioa, Spain}
\affiliation{IKERBASQUE, Basque Foundation for Science, E-48011 Bilbao, Spain}
\author{A.\ Matzkin}
\affiliation{Laboratoire de Physique Th\'eorique et Mod\'elisation, CNRS Unit\'e 8089, CY
Cergy Paris Universit\'e, 95302 Cergy-Pontoise cedex, France}

\begin{abstract}
We investigate wavepacket dynamics across supercritical barriers for the
Klein-Gordon and Dirac equations. Our treatment is based on a multiple
scattering expansion (MSE). For spin-0 particles, the MSE diverges, rendering
invalid the use of the usual connection formulas for the scattering basis
functions.\ In a time-dependent formulation, the divergent character of the
MSE naturally accounts for charge creation at the barrier boundaries. In the
Dirac case, the MSE converges and no charge is created. We show that this
time-dependent charge behavior dynamics can adequately explain the\ Klein
paradox in a first quantized setting. We further compare our semi-analytical
wavepacket approach to exact finite-difference solutions of the relativistic
wave equations.

\end{abstract}
\pagenumbering{gobble} 
\maketitle

%\clearpage
\pagenumbering{arabic} 
\setcounter{page}{1}

\section{Introduction}

One of the salient features of relativistic quantum mechanics is the intrinsic
mixture of particles with their corresponding antiparticles.\ This aspect
already appears in "first quantized" single particle Relativistic Quantum
Mechanics (RQM) \cite{greiner}. This phenomenon becomes prominent when a
particle is placed in a very strong external classical field, a field strong
enough to close the gap between particles and antiparticles -- a
supercritical field.

Most textbooks, as well as the overwhelming majority of past works employ
time-independent stationary phase or plane wave arguments when dealing with
supercritical fields within RQM. As is well known, in quantum scattering
theory time-independent quantities such as cross-sections or energy levels can
be readily computed, but it is often ambiguous to attempt to infer the
dynamics from considerations involving a single plane wave. This is even more
the case when relativistic phenomena are investigated: one has to deal with
additional difficulties, such as the breakdown of the usual connection
formulas linking the scattering solutions in different regions, or the Klein
paradox -- a phenomenon usually expressed as a current reflected from a
supercritical step or barrier higher than the incident one. Standard textbooks
(e.g. \cite{bjorken,greiner,strange,wachter}) give different, often
conflicting accounts of the Klein paradox. This situation is reflected even in
recent works
\cite{DiracSpin,kim2019,morales19,xue,no-klein-tunnel-dirac,noKP-Dirac,bosanac,emilio,leo2006,noKP-KG,nitta}%
, that reach different conclusions generally based on time independent considerations.

In part for this reason, it is often stated, in different situations dealing
with the Klein paradox, that a RQM approach is inadequate and that a
quantum field theory (QFT) treatment is necessary (see e.g. \cite{strange} for the step case, or
\cite{xue} for the barrier case). Nevertheless even in a first quantized
framework the RQM wave equations have charge creation built-in. We argue in
this paper that this aspect is best understood by considering the
time-dependent wavefunction dynamics. We will assess whether this leads to a
consistent first quantized explanation of the Klein paradox. Interestingly,
other very recent works have focused on different time-dependent aspects of
the relativistic wave equations \cite{transient,KG-moving}.

In this paper we will develop a time-dependent wavepacket treatment suited to
investigate the dynamics of spin-0 bosons and spin-1/2 fermions in model
supercriticial barriers. In order to develop our semi-analytic wavepacket
approach, we will rely on a Multiple Scattering Expansion (MSE).\ We will see
that the nature of the MSE is different for solutions of the Klein-Gordon and
Dirac equations. In the Klein-Gordon case, the MSE diverges, physically
corresponding to charge creation as the wavepacket hits a barrier's edge. This
implies that the usual connection formulas between wavefunctions in different
regions -- which are obeyed when a single step potential is considered -- are
not applicable when dealing with supercritical barriers. Employing connection
formulas leads to inconsistencies, like superluminal traversal times, that
were recently noted \cite{xue} though (incorrectly, as we will see) attributed
to the limitations of the first quantized formalism. In the Dirac case, the
MSE converges, as no charge is created when the wavepacket hits the barrier.
This leads to an interpretation of Klein tunneling that is qualitatively
different from the bosonic case. This also ensures that the connection
formulas, that have been employed in a countless number of works, remain valid.

The multiple scattering expansion will be given in Sec.\ \ref{sec2}, after
introducing the model barriers we will be working with. These ingredients will be employed
in Sec. \ref{sec3} to build wavepackets. We will then give a couple of examples displaying the dynamics of
a bosonic or fermionic particle impinging on a supercritical barrier. Our
wavepacket results will be compared to numerical solutions obtained from a
code we have developed to solve numerically the relativistic wave equations.
Our results will be discussed in Sec. \ref{sec4}. We will more specifically
focus on the extent to which the Klein paradox can be accounted for within a
first quantized framework. We close the paper by a Conclusion.

\section{Charge creation and Multiple Scattering Expansions \label{sec2}}

\subsection{Potential barriers}

We are interested in this work by the dynamics of a relativistic
\textquotedblleft particle\textquotedblright\ impinging on a one dimensional static
barrier of width $L$. The barrier should be discriminated from the step, which is by far the case that has most
often been considered in studies of the Klein paradox. The relevant wave equations for spin-0 and spin-1/2 particles are
recalled in Appdx A. 

The simplest case is the rectangular barrier defined by $V(x)=V\theta(x)\theta(L-x)$
%\begin{equation}
%V(x)=V\theta(x)\theta(L-x) \label{rb}%
%\end{equation}
 where $\theta$ denotes the unit-step function and $V$ denotes the barrier
height. We will also consider smooth barriers, for which we will employ the
potential $V_{s}(x,\epsilon)=\frac{V}{2}\left[  \tanh(\epsilon x)-\tanh\left(
\epsilon(x-L)\right)  \right] $
%\begin{equation}
%V_{s}(x,\epsilon)=\frac{V}{2}\left[  \tanh(\epsilon x)-\tanh\left(
%\epsilon(x-L)\right)  \right]  \label{sv}%
%\end{equation}
 since for this potential analytical solutions are known \cite{kennedy};
$\epsilon$ is the smoothness parameter.

\subsubsection{Subcritical barriers}

Let us consider a rectangular barrier. Plane wave
solutions of the canonical KG equation (see Appdx A) are of the form%
\begin{equation}
\psi_{j}^{\pm}(t,x)=\left(  A_{j}^{\pm}e^{ip_{j}x/\hbar}+B_{j}^{\pm}%
e^{-ip_{j}x/\hbar}\right)  e^{\mp iE(p_{1})t/\hbar}\label{pw}%
\end{equation}
where $j=1,2,3$ denotes the regions depicted in Fig. \ref{figBarrier} and the
$\pm$ signs corresponds to states with positive and negative energies $\pm
E(p_{1})$ with $E(p_{1})=\sqrt{m^{2}c^{4}+p_{1}^{2}c^{2}}$. $p_{j}$ is the
momentum; for positive energies, $p_{j}>0$ gives a wave moving from left to
right (but from right to left for negative energies).

As is well known \cite{manogue,greiner}, for \textquotedblleft
subcritical\textquotedblright\ potentials ($p_{2}$ is imaginary) plane-waves
scattering is similar to the usual non-relativistic situation (small
transmission amplitude and exponentially decreasing waves). Assume boundary
conditions for which an incident positive energy wave travels from left to
right; this imposes $B_{3}=0$ and for definiteness we set the incoming
amplitude to $A_{1}=1$. The other amplitudes $A_{j}$ and $B_{j}$ are deduced
by matching the wavefunctions and their space derivatives at the boundaries
$x=0$ and $x=L$ (for reasons that will become clear below, we will not need to
deal with boundary conditions for negative energy plane waves; we henceforth
write $A$ for $A^{+},$ etc.). This way of obtaining the amplitudes does not
necessarily hold when $V$ becomes supercritical.

\subsubsection{Supercritical barriers}

A supercritical potential is a potential high enough to give rise to Klein
tunneling \cite{review99}, whereby the incoming wavepacket penetrates undamped
($p_{2}$ is real) inside the barrier.  In the bosonic case,
this gives rise \cite{manogue} to superradiance (a reflected current higher
than the incoming one). In the fermionic case there is no superradiance
(although some authors suggest differently, see e.g. \cite{bjorken,leo2006}),
and supercritical steps have been deemed to have an acceptable interpretation
only within a QFT approach
\cite{hansen,grobe-review,gavrilov}, a point that appears to be supported by
the wide variety of conflicting interpretations of Klein tunneling that have
been proposed within the first quantized framework
\cite{DiracSpin,kim2019,morales19,xue,no-klein-tunnel-dirac,noKP-Dirac,bosanac,emilio,leo2006,noKP-KG,nitta}%
.

\subsection{Multiple Scattering Expansions \label{MSEsec}}

\begin{figure}[tb]
	\includegraphics[width=4.5cm]{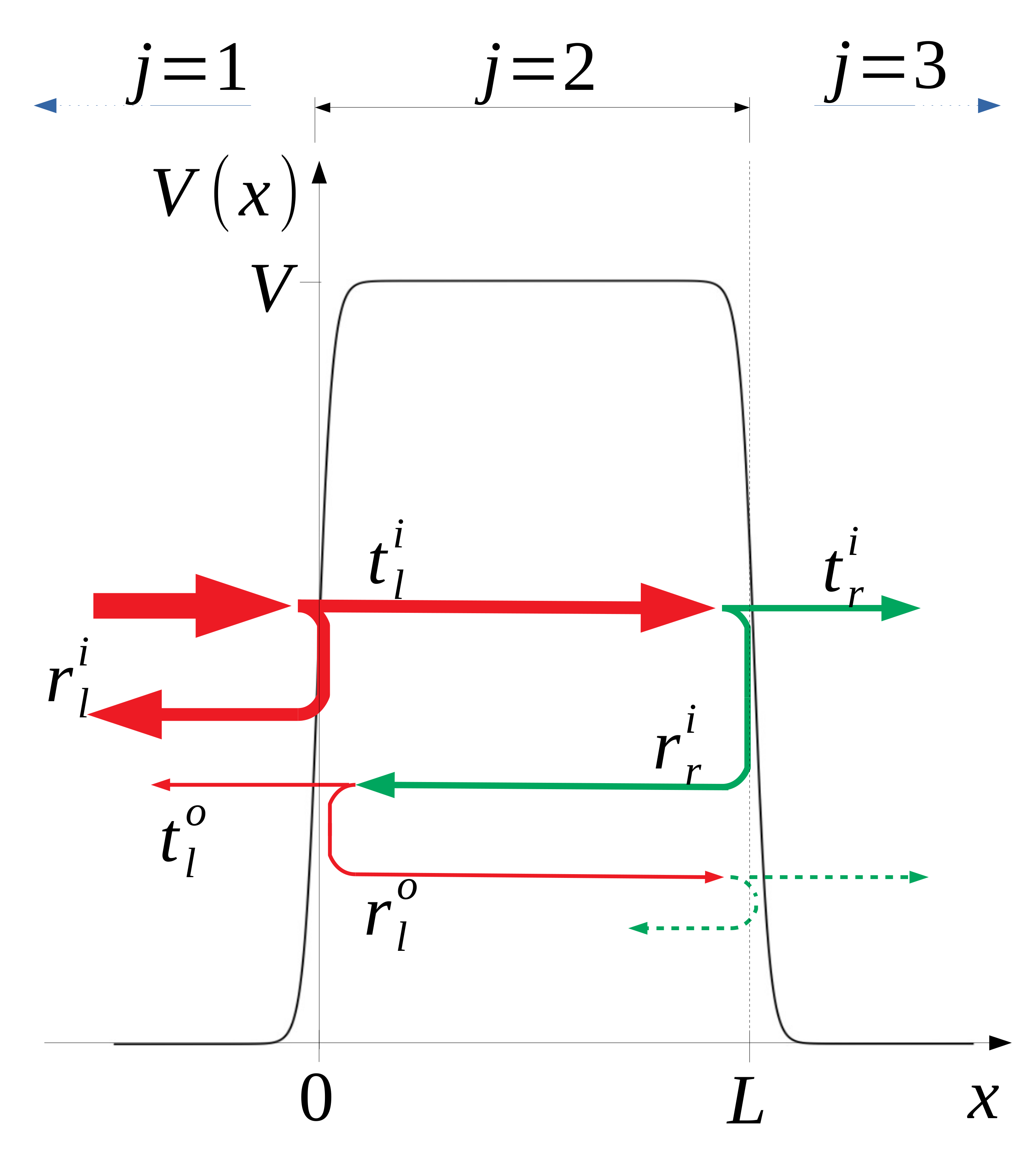} \caption{A generic barrier is shown
		along with the $3$ regions $j=1,2,3$. The scattering amplitudes of the
		Multiple Scattering Expansion at the left ($l$) and right ($r$) edges are
		indicated for an incoming wave boundary condition (see text for details). }%
	\label{figBarrier}%
\end{figure}

It is well-known that when transmission of waves across several media takes
place, one has to take into account a multiple scattering process.\ Referring
again to Fig. \ref{figBarrier}, consider an asymptotically free (at
$x=-\infty$) wave coming toward the barrier (we set again $A_{1}=1$ and
$B_{3}=0$). Reflection of the incoming wave on the barrier takes place with
amplitude $r_{l}^{i}$, which is the reflection amplitude of a step. The
transmitted amplitude at that point, $t_{l}^{i}$ is that of a step, but the
wave traveling inside the barrier reaches the right edge and gets transmitted
and reflected with amplitudes $t_{r}^{i}$ and $r_{r}^{i}$. This reflected
wave travels back towards the left side of the barrier, getting reflected and
transmitted with coefficients $r_{l}^{o}$ and $t_{l}^{o}$. This process is
iterated an infinite number of times yielding%
\begin{equation}%
\begin{tabular}
[c]{ll}%
$r\equiv B_{1}=r_{l}^{i}+t_{l}^{i}t_{l}^{o}r_{r}^{i}\sum_{n\geq0}(r_{l}%
^{o}r_{r}^{i})^{n}\quad$ & $t\equiv A_{3}=\sum_{n\geq0}t_{l}^{i}\left(
r_{l}^{o}r_{r}^{i}\right)  ^{n}t_{r}^{i}$\\
$A_{2}=\sum_{n\geq0}t_{l}^{i}\left(  r_{l}^{i}r_{l}^{o}\right)  ^{n}$ &
$B_{2}=\sum_{n\geq0}t_{l}^{i}r_{r}^{i}\left(  r_{l}^{o}r_{r}^{i}\right)  ^{n}%
$\\
$A_{1}=1$ & $B_{3}=0.$%
\end{tabular}
\ \ \ \label{iter}%
\end{equation}
The amplitudes obtained by using this Multiple Scattering Expansion (MSE)
should match those obtained by employing the usual connection formulas at the
boundaries, but this will happen only provided the sums in Eq. (\ref{iter}) converge.

\subsubsection{Klein-Gordon equation: Divergent Multiple Scattering Expansion
\label{KGsec}}

Let us consider the KG equation in the presence of a supercritical step
$V\theta(x)$ at $x=0$. Scattering of a positive energy plane wave coming from
the left, as given by Eq. (\ref{pw}) sets $\bar{A}_{1}=1,\bar{B}_{2}=0$ (we
use the bar to avoid confusion between the step and barrier amplitudes).
$\bar{A}_{2}$ and $\bar{B}_{1}$ are thus obtained by applying the boundary
conditions at $x=0$, yielding $\bar{B}_{1}=(p_{1}-p_{2})/(p_{1}+p_{2})$ and 
$\bar{A}_{2}=2p_{1}/(p_{1}+p_{2})$
%\begin{equation}
%\bar{B}_{1}=\frac{p_{1}-p_{2}}{p_{1}+p_{2}}\qquad\bar{A}_{2}=\frac{2p_{1}%
%}{p_{1}+p_{2}},
%\end{equation}
where following our notation given above, $p_{1}>0$ is the momentum of the
incoming plane wave (in region 1) and
\begin{equation}
p_{2}(p_{1})=-\frac{1}{c}\sqrt{p_{1}^{2}c^{2}-2V\left(  m^{2}c^{4}+p_{1}%
^{2}c^{2}\right)  ^{1/2}+V^{2}}.\label{KGp2}%
\end{equation}

The amplitudes $\bar{B}_{1}$ and $\bar{A}_{2}$ of the step correspond to the
amplitudes $r_{l}^{i}$ and $t_{l}^{i}$ of the barrier MSE, Eq. (\ref{iter}).
$r_{r}^{i}$ and $t_{r}^{i}$ are obtained by considering the step
$V\theta\left(  L-x\right)  $, and $r_{l}^{o}$ and $t_{l}^{o}$ arise from the
step $V\theta(x)$ with a wave coming from the right (see Appdx B). By inserting the values for these elementary scattering amplitudes
into the MSE of Eq. (\ref{iter}), we obtain the scattering amplitudes for the
rectangular barrier. These amplitudes converge if $|(p_{1}-p_{2})/(p_{1}+p_{2})|<1$ %
%\begin{equation}
%\left(  \frac{p_{1}-p_{2}}{p_{1}+p_{2}}\right)  ^{2}<1\label{conv}%
%\end{equation}
 in which case it can be checked that the amplitudes $A_{j}$ and $B_{j}$ match
the ones obtained by using the connection formulas at the boundaries.

This condition is fulfilled for subcritical barriers (both $p_{1}$
and $p_{2}$ are positive) but for supercritical barriers, we have $p_{1}>0$
and $p_{2}<0$ and the MSE\ diverges. This divergence does not make much sense
in a stationary plane-wave picture, in which the scattering amplitudes become
infinite, but we will see below in Sec. \ref{sec3} that in a time-dependent
approach, the divergence corresponds physically to the creation of charge each
time a wavepacket hits a barrier edge. 

Interestingly, if the ``converged'' amplitudes (usually obtained by employing
the connection formulas) are employed in the supercritical case, unphysical
results are obtained. It was recently noticed \cite{xue} in a plane wave
analysis that the barrier traversal time defined from the phase energy
derivative was superluminal in the supercritical case, an unphysical result
attributed to the limitations of the first quantized formalism. In a
wavepacket approach, building wavepackets with the converged amplitudes
results in an acausal wavepacket coming out from the right side of the
supercritical barrier before the incoming wavepacket has even hit the barrier
\cite{paper1}. Other works have also employed the connection formulas in a
supercritical context \cite{kim2019,emilio}.

\subsubsection{Dirac equation: Convergent Multiple Scattering Expansion}

The derivation of the MSE for the Dirac equation is similar to the one we have
given for the KG\ equation. The plane wave solutions of the 2-component state
$\Phi(t,x)$ take the form:%
\begin{equation}
\Phi_{j}^{\pm}(t,x)=\left[  A_{j}^{\pm}\left(
\begin{array}
[c]{c}%
1\\
\alpha_{j}^{\pm}(p_{1})
\end{array}
\right)  e^{^{ip_{j}x/\hbar}}+B_{j}^{\pm}\left(
\begin{array}
[c]{c}%
1\\
-\alpha_{j}^{\pm}(p_{1})
\end{array}
\right)  e^{^{-ip_{j}x/\hbar}}\right]  e^{-i\sqrt{m^{2}c^{4}+p_{1}^{2}c^{2}%
}t/\hbar}n_{j}^{\pm}(p_{1})\label{Dpw}%
\end{equation}
where the coefficients $\alpha_{j}^{\pm}$ are given by Eqs. (C-5) to (C-8) of Appdx C.$j$ refers again to
the 3\ regions depicted in Fig.\ \ref{figBarrier} and $n_{j}^{\pm}$ are
normalization coefficients.

As in the KG case, let us treat the barrier as a multiple scattering
expansion, with the same boundary conditions.
The amplitudes $A_{j}^{\pm}$ and $B_{j}^{\pm}$ are again given by Eq.
(\ref{iter}), with coefficients $r_{l}^{i},$ $t_{l}^{i}...$ that are different
for positive and negative energy wavefunctions. As in the KG\ case, the MSE
for a rectangular barrier is built from the reflection and transmission
amplitudes $\bar{A}_{j}^{\pm}$ and $\bar{B}_{j}^{\pm}$ of the steps
$V\theta(x)$ and $V\theta(L-x)$. The matching condition $\Phi_{1}%
^{+}(t,x=0)=\Phi_{2}^{+}(t,x=0)$, yields%
\begin{equation}
\bar{B}_{1}^{+}=\frac{\alpha_{1}^{+}(p_{1})-\alpha_{2}^{+}(p_{1})}{\alpha
_{1}^{+}(p_{1})+\alpha_{2}^{+}(p_{1})}\qquad\bar{A}_{2}^{+}=\frac{2\alpha
_{1}^{+}(p_{1})}{\alpha_{1}^{+}(p_{1})+\alpha_{2}^{+}(p_{1})}\frac{n_{1}^{+}%
}{n_{2}^{+}}.
\end{equation}
The amplitudes $\bar{B}_{1}$ and $\bar{A}_{2}$ of the step correspond to the
amplitudes $r_{l}^{i+}$ and $t_{l}^{i+}$ entering the MSE for the barrier. The
other step amplitudes are obtained in the same manner, see Eq. (B-5) in the Appdx. It can be seen that the series
converge provided $\lvert\frac{\alpha_{1}^{+}(p_{1})-\alpha_{2}^{+}(p_{1})}{\alpha_{1}^{+}%
	(p_{1})+\alpha_{2}^{+}(p_{1})}\rvert<1$
%\begin{equation}
%\lvert\frac{\alpha_{1}^{+}(p_{1})-\alpha_{2}^{+}(p_{1})}{\alpha_{1}^{+}%
%(p_{1})+\alpha_{2}^{+}(p_{1})}\rvert<1\label{Dcond}%
%\end{equation}
true since $\alpha_{2}^{+}(p_{1})$ is positive when $V$ is supercritical.

Therefore for the Dirac equation, the MSE converges.\ The usual connection
formulas at $x=0$ and $x=L$ may be employed to obtain directly the barrier
amplitudes $A_{j}^{\pm}$ and $B_{j}^{\pm}$ given by Eqs. (C-9) to (C-12). Note that most past works (e.g.,
\cite{emilio,graphene,barrierDex1}) have indeed employed such connection
formulas without however examining the justifications for their use.

\section{Wavepacket dynamics \label{sec3}}

\subsection{Construction from plane-wave expansions}

The most straightforward way to construct wavepackets starting from an initial
distribution is to employ a plane-wave expansion, valid everywhere except in the slope region
for a sufficiently steep barrier.

\subsubsection{Klein-Gordon equation wavepackets}

The Klein-Gordon plane-waves were given in Eq. (\ref{pw}). These solutions can
be expressed in Hamiltonian form (see Appdx A), as
\begin{equation}
\Psi_{j}^{\pm}(t,x)=N%
\begin{pmatrix}
mc^{2}\pm\sqrt{m^{2}c^{4}+p_{j}^{2}c^{2}}\\
mc^{2}\mp\sqrt{m^{2}c^{4}+p_{j}^{2}c^{2}}%
\end{pmatrix}
\left(  A_{j}^{\pm}e^{ip_{j}x/\hbar}+B_{j}^{\pm}e^{-ip_{j}x/\hbar}\right)
e^{\mp i\sqrt{m^{2}c^{4}+p_{1}^{2}c^{2}}t/\hbar}\ ,\label{pwham}%
\end{equation}
where $p_{3}=p_{1}$ and $p_{2}$ is given by Eq. (\ref{KGp2}), the amplitudes
$A_{j}^{\pm}$ and $B_{j}^{\pm}$ are given by Eq.\ (\ref{iter}), and $N$ is a
global normalisation constant.

Assume that at $t=0$ we have an initial wavefunction $G(t=0,x)=\left(
\varphi_{G}(0,x),\chi_{G}(0,x)\right)  $ in region 1, to the left of the
barrier. The time evolution can be employed by applying the pseudo-unitary
evolution operator on $G(t=0,x),$ or equivalently by using the Fourier
transform $G(t=0,x)=\int dpe^{ipx/\hbar}\hat{G}(t=0,p).$ \ The time
evolved wavepacket can then be written as
\begin{equation}
G(t,x)=\sum_{j}\theta_{j}\int dp_{1}\left(  c_{KG}^{+}(p_{1})\Psi_{j}%
^{+}(t,x;p_{1})+c_{KG}^{-}(p_{1})\Psi_{j}^{-}(t,x;p_{1})\right)
,\label{KGwp-g}%
\end{equation}
where $\theta_{j}$ ensures each expression is used only in the region $j$ in
which it is valid, as per Fig. \ref{figBarrier} (explicitly, $\theta
_{1}=\theta(-x),\theta_{2}=\theta(x)\theta(L-x)$ and $\theta_{3}=\theta(x-L)$).

\begin{figure}[tb]
	\includegraphics[width=9.4cm]{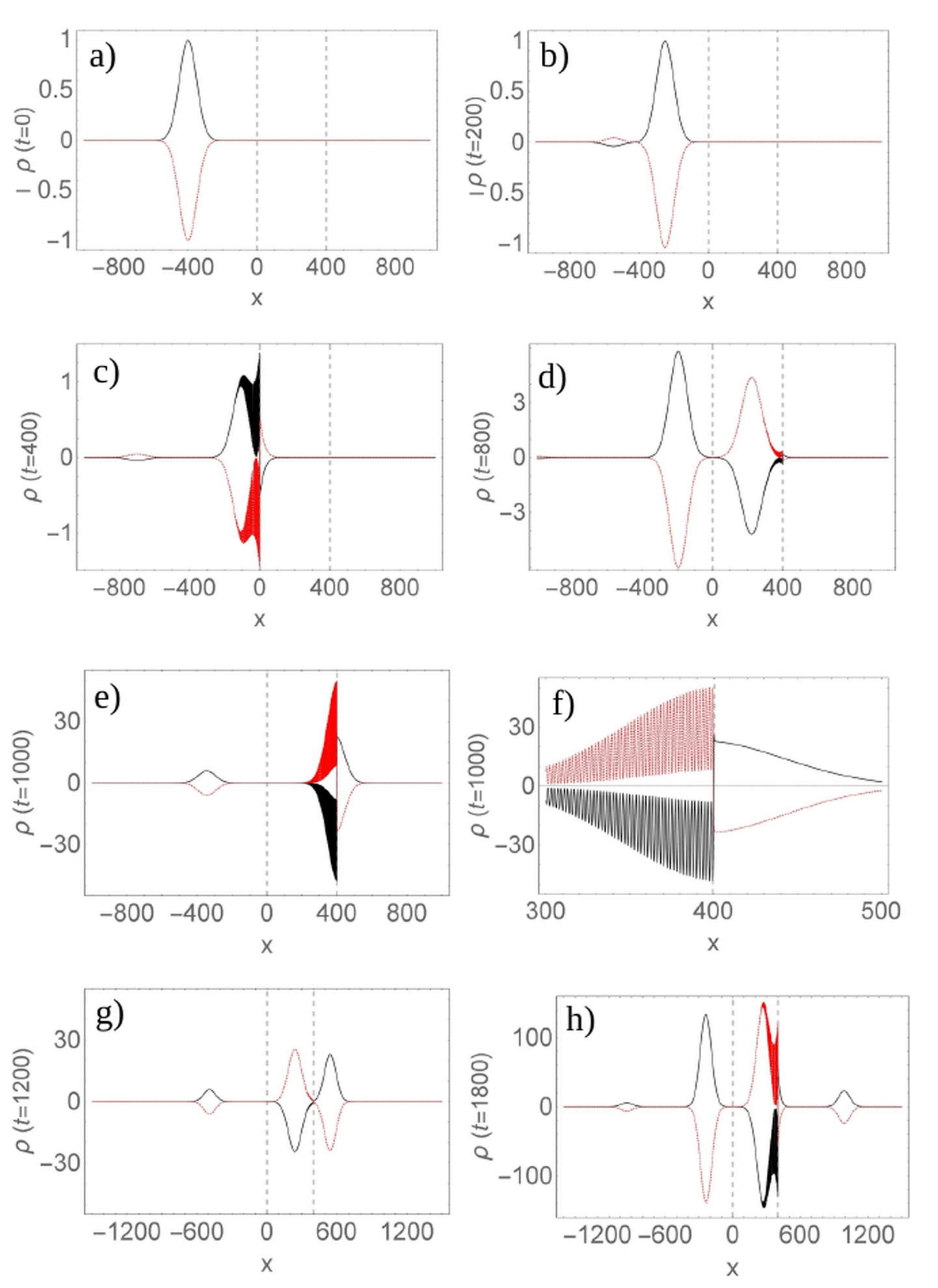} \caption{Wavepacket dynamics for
		a spin-0 boson described by the Klein Gordon equation impinging on a smooth
		supercritical barrier. The charge density $\rho(t)$ is given for different
		times as specified on each plot. Our semi-analytic wavepacket approach is
		shown in black solid lines, while for the sake of comparison our
		finite-difference solutions are shown upside-down with a dashed (online: red)
		line. The supercritical barrier lies within the dotted vertical gray lines.
		The same initial state, shown in a) is taken for the wavepacket and the
		numerical calculations and no adjustment or renormalization are made at longer
		times (the change in scale reflects charge creation). Note that f) is a zoom
		of e) in the region around the right edge of the barrier. The values of the
		parameters employed are given in the main text. Natural units ($\hbar
		=c=m=\varepsilon_{0}=1$) are used \cite{units}.}%
	\label{fig-kg}%
\end{figure}

To be specific, we will choose an initially localized state of the form%
\begin{equation}
G(0,x)=(\exp(\frac{-(x-x_{0})^{2}}{4d^{2}}-ip_{0}x/\hbar),0)%
%\begin{pmatrix}
%\exp(\frac{-(x-x_{0})^{2}}{4d^{2}})\\
%0
%\end{pmatrix}
%e^{-ip_{0}x/\hbar}.
\label{is1}%
\end{equation}
Picking $x_{0}$ far to the left of the barrier and $p_{0}>0$, the choice
(\ref{is1}) gives an initial state with positive charge. The coefficients
$c_{KG}^{\pm}=\left\langle \Psi_{1}^{\pm}\right\vert \left.
G\right\rangle _{KG}$ are readily computed [Eq. (C-4)] and it can be seen
that $c_{KG}^{-}$ is non-zero (although it is small for non-relativistic
velocities). This negative energy component moves to the left (recall that
$p_{0}>0$ yields an antiparticle that moves in the negative direction), so
only the positive energy wavepacket (particle) impinges on the barrier
\footnote{We might as well have chosen to restrict $G(0,x)$ to the expansion
over the positive energy eigenstates in Eq. (\ref{KGwp-g}), but when comparing
with numerical solutions it is more interesting to keep (\ref{is1}) as the
initial state.}. 

\subsubsection{Dirac equation wavepackets}

A similar construction can be used to build wavepackets evolving according to
the Dirac equation, starting from an initial state $\left\vert
G(t=0)\right\rangle $ of the same form as Eq. (\ref{is1}) now expanded over
the Dirac equation plane-wave solutions. The time evolved wavepacket is%
\begin{equation}
G(t,x)=\sum_{j}\theta_{j}\int dp_{1}\left(  c_{D}^{+}(p_{1})\Phi_{j}%
^{+}(t,x;p_{1})+c_{D}^{-}(p_{1})\Phi_{j}^{-}(t,x;p_{1})\right)  ,\label{D-wp}%
\end{equation}
where the  coefficients $c_{D}^{\pm}=\left\langle \Phi_{1}^{\pm}\right\vert
\left.  G\right\rangle $ are given by Eq. (C-13).

\subsection{Numerical solutions}

We have computed numerical solutions to the Klein-Gordon and Dirac equations.
This was done by discretizing the corresponding evolution operator in real
space for small time steps. The initial wavefunction $G(t=0,x)$ is discretized
on a fixed space-grid and the derivatives in the evolution operator are
approximated by finite differences in the fourth or fifth-order
approximation.\ The computational details are given in Appdx D (see also Refs. \cite{keitel,bauke,muller}).

\subsection{Illustrative Results}

We will now illustrate our wavepacket approach  and compare it with fully
numerical solutions obtained by solving the relativistic wave equations with a
finite-difference scheme. 

Fig. \ref{fig-kg} illustrates the wavepacket dynamics for a spin-0\ boson
impinging on a smooth barrier $V_{s}(x,\epsilon)$ with $\epsilon=5,$ $L=400$
and $V=3.4$ (we use natural units \cite{units}). We pick the initial state Eq.
(\ref{is1}) with $x_{0}=-400,$ $p_{0}=\sqrt{5}/2$ and $d=50$ (in order to have
a rather narrow momentum distribution). We also provide numerical solutions
obtained by using our finite-difference scheme introduced above (these
solutions are plotted upside-down). The wavepacket moves towards the barrier
(save for an antiparticle component moving to the left, visible at $t=200$),
and has appreciably hit the barrier by $t=400$, while at $t=800$ one sees
Klein tunneling accompanied by charge production both inside and outside the
left edge of the barrier (note the vertical scale). At $t=1000$, the
antiparticle wavepacket hits the right edge of the barrier, inducing
additional charge production both for the transmitted (particle) wavepacket
and for the reflected (antiparticle) one. This motion continues, with the
amplitude inside the barrier growing at each reflection.

The MSE\ based wavepacket dynamics match very well the computations obtained
from the finite-difference solutions. In practice, the number of terms that
need to be taken into account in the MSE sum (\ref{iter}) is congruent with
the time $t$ at which the wavepacket is computed. Indeed, the $n$th term
corresponds formally to a wavepacket translated by the order of $2n\pi L$,
that for values of $n$ that are high enough did not have time to reach the
barrier. Hence these terms do not contribute to the wavepacket (in the
calculations shown on Fig. \ref{fig-kg}, including terms up to $n=4$ is
sufficient). 

\begin{figure}[h]
\includegraphics[width=9cm]{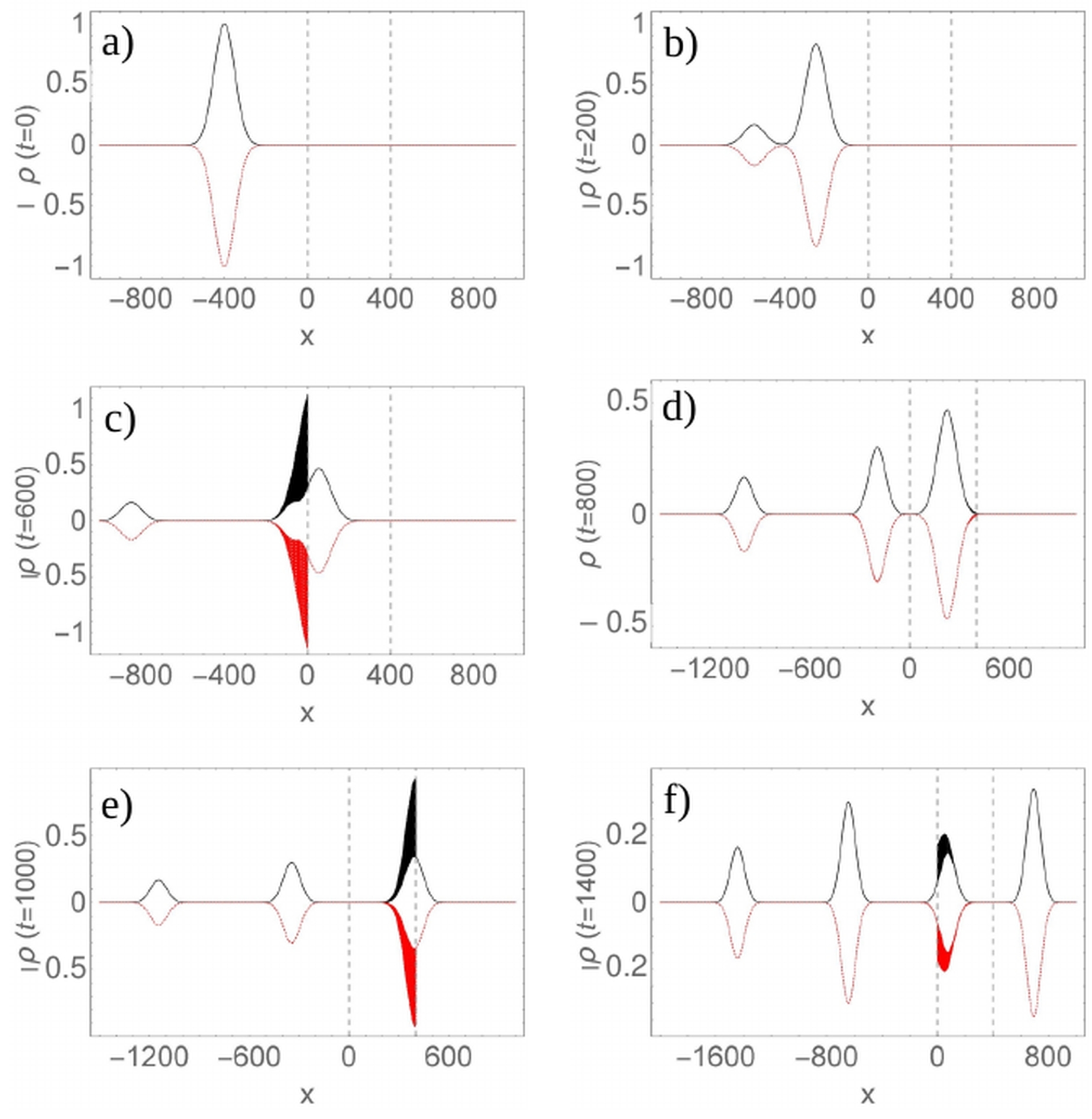} \caption{Wavepacket dynamics for
a spin-1/2 fermion described by the Dirac equation impinging on a
supercritical barrier. The density $\rho(t)$ is given for different times as
specified. The semi-analytic wavepacket approach gives the results shown in
black solid lines, while the finite-difference solutions are shown upside-down
with a dashed (online: red) line. The supercritical barrier lies within the
dotted vertical gray lines. The same initial state, shown in a) is taken for
the wavepacket and the numerical calculations and no adjustment or
renormalization are made at longer times. The values of the parameters
employed are given in the main text. Natural units ($\hbar=m=c=\varepsilon
_{0}=1$, where $m$ is the electron mass) are used.}%
\label{fig-D}%
\end{figure}

The wavepacket dynamics for a spin 1/2 fermion is shown in Fig. \ref{fig-D}%
.\ We picked the same barrier and initial state parameters as in the
Klein-Gordon case except that the smoothness parameter was taken to be the
size of the space integration step in the numerical code ($\epsilon\simeq
250$), effectively corresponding to the rectangular barrier limit.\ We note
again the very good agreement between the wavepackets constructed with Eq.
(\ref{D-wp}) and the numerical solutions.

\section{Discussion\label{sec4}}

\subsection{General remarks}

We have developed a wave-packet approach to describe Klein tunneling across a
supercritical barrier both for Klein-Gordon and Dirac particles. The main
ingredient was seen to be the multiple scattering expansion, that diverges in
the KG\ case but converges in the Dirac case.\ 

The main issue now is whether these results lend to a consistent
interpretation of the Klein paradox within the first quantized
formalism.\ This is known to be problematic when basing considerations on
stationary plane waves, and indeed different, sometimes conflicting
interpretations of superradiance and supercritical tunneling in a first
quantized framework have been given
\cite{DiracSpin,kim2019,morales19,xue,no-klein-tunnel-dirac,noKP-Dirac,bosanac,emilio,leo2006,noKP-KG,nitta}%
. Klein tunneling has even been denied to exist in the bosonic \ case
\cite{noKP-KG} or for fermions \cite{noKP-Dirac}. Of course since particle
creation is induced by supercritical potentials, a proper approach requires a
QFT treatment. But time-dependent QFT approaches to
tackle this problem are scarce (with the exception of the work reviewed in Ref
\cite{grobe-review}, where supercriticial steps, rather than barriers, were
investigated both for fermions and bosons; see also \cite{grobe-recent} and
Refs.\ therein for recent ramifications of this \ time-dependent QFT method).
Hence, although time-independent approaches to Klein tunneling within the
first quantized framework might be seen as unreliable, it would be worthwhile
to consider the merits of a time-dependent first quantized account, and
examine its consistency with time-dependent QFT approaches.

\subsection{Klein Paradox in a first quantized framework}

\subsubsection{Bosons}

For the Klein-Gordon equation, the account seems rather simple: a
supercritical potential creates positive and negative charges in equal amounts
but in different spatial regions. In our wavepacket approach, this is
explained by the divergent character of the multiple scattering expansion,
although this property is encapsulated in the pseudo-unitary evolution
operator that is solved numerically in the finite-difference code.

Hence, when a particle impinges on a supercritical barrier, the reflected
wavepacket has a higher amplitude than the incoming one. Similarly, when the
Klein tunneling wavepacket reaches the right edge of the barrier, a positively
charged wavepacket leaves the barrier, and this is compensated by additional
negative charge inside the barrier. This process goes on indefinitely with
positive and negative charge increasing at each reflection (to the extent that
the supercritical potential can be maintained despite the growing charge
inside the barrier).

It is difficult to accommodate the charge creation mechanism by supercritical
fields with the idea that the first quantized framework would describe a
single particle in a superposition of states of different charges (rather than
explaining this feature in terms of particle-antiparticle creation). However,
leaving this important physical issue aside, we see that the time-dependent
wavepackets by themselves do account for the bosonic Klein paradox.

\subsubsection{Fermions}

For the Dirac equation, the dynamics is very different. Assume the incoming
particle is an electron (with negative charge). Part of the incoming
wavepacket gets reflected by the barrier, and part gets transmitted.\ Contrary
to the bosonic case, there is no charge creation, so
the reflected wavepacket has a smaller amplitude than the incoming one.\ Since
the total probability density is conserved, requiring charge conservation
implies that the wavepacket tunneling inside the barrier also has negative
charge. This is difficult to accommodate within a single particle picture,
even by relying on hole theory.

Indeed, according to the standard hole theory account of the Klein paradox for
fermions (see Ch. 12 of \cite{greiner}), the energy levels of the Dirac sea
are raised inside the barrier by the supercritical potential. The incoming
electron can then \textquotedblleft knock off\textquotedblright\ an electron
from the Dirac sea inside the barrier. This would account for the reflected
electron, and would leave a hole in the Dirac sea, corresponding to a
positively charged positron. However we have seen the wavefunction we have
obtained inside the barrier has an overall negative charge, so that in
addition to the hole an electron should also propagate inside the barrier.

\subsection{Relation to time-dependent QFT}

Klein tunneling and the Klein paradox ultimately depend on a multiparticle
process involving particle-antiparticle pairs creation, that should therefore
be described by QFT. A time-dependent QFT scheme \cite{grobe-review} has
investigated Klein tunneling for rectangular and smooth steps.\ The QFT
calculations have been compared to numerical solutions of the Klein-Gordon
\cite{boson-qft-grobe} and Dirac \cite{dirac-qft-grobe} equations.

\subsubsection{Bosonic QFT}

In the absence of any incoming particle, space-time dependent charge densities
obtained from the bosonic field operator indicate that the supercritical step
induces pair creation, particles with positive energies moving away from the
step and wavepackets with negative energies (antiparticles) ``tunneling''
inside the step \cite{boson-qft-grobe}. When a particle is sent towards the
barrier, the QFT computations show that pair creation is enhanced (precisely
by the transmission amplitude $t_{l}^{i}$ of Eq. (\ref{iter})). This
enhancement corresponds to the first quantized computations for the step. The
same correspondence between QFT and our results can be thought to hold in the
barrier case: the wavepackets we obtained would correspond to the pair
creation enhancement produced by sending a particle on the barrier, on top of
the pair creation process out of the vacuum. Note that in the barrier case,
spontaneous pair creations are also expected to lead to an amplification
mechanism through multiple reflections inside the barrier.

\subsubsection{Fermionic QFT}

In the Dirac case, QFT computations of the time-dependent spatial densities
from the fermionic operators for a step show that
an incoming electron modifies the pair production process induced by the
supercritical field \cite{dirac-qft-grobe}. The reflected fraction of the
incoming first quantized wavepacket appears as an excess of the particle
charge (relative to the charge produced by the step). The transmitted
wavepacket, propagating inside the step, appears instead as a dip in the
anti-particle (positronic) charge produced by the supercritical potential. The
interpretation is that, as a result of the Pauli principle, the incoming
electron decreases the pair-production process that takes place at the
step.\ The decrease of antiparticle production at the edge of the step gives
rise to a hole in the positronic charge inside the step. The dip in the
electron production is partially compensated by the incoming electron that is
interpreted \cite{dirac-qft-grobe} as being fully reflected, yielding overall
an excess of electronic charge which is the reflected first quantized wavepacket.

It is not obvious whether one can extrapolate straightforwardly the QFT
results for a step to a fermionic barrier. Now both edges of the barrier have
a pair production process. Inside the barrier, the Pauli principle also
applies to positrons. This has no counterpart in a first quantized framework.
Nevertheless, one can speculate that, at least for a sufficiently wide
barrier, an incoming electron partially suppresses pair production (similarly
to the step) at the left edge of the barrier. The hole in positronic charge
propagating inside the barrier partially lifts the blockade due to the Pauli
principle. Upon reaching the right edge of the barrier, this results in an
enhancement in pair production. The wavepacket coming out from the barrier in
our first quantized calculations should therefore correspond to the additional
electrons produced on the right side of the barrier as the hole reaches the
right edge. The additional positrons that are produced are responsible for the
smaller amplitude of the hole reflected inside the barrier. This process
continues inside the barrier with the hole oscillating with decreasing
amplitude. While this QFT-based picture, hinging on the results obtained for a
step appear to be consistent with the present first-quantized results,
time-dependent QFT computations for the barrier would be needed in order to
confirm the precise relationship between both pictures of supercritical Klein tunneling.

\section{Conclusion\label{conc}}

We have investigated in this paper the Klein paradox for barriers from a
time-dependent perspective within the first quantized framework. We have
developed a semi-analytical wavepacket approach relying on the properties of a
multiple scattering process inside the barrier. This yields a very different
behavior for bosons and fermions. In the bosonic case, each collision of the
wavepacket on an edge of the supercritical potential creates charge (as the
multiple scattering process diverges), leading to superradiance. In the Dirac
case Klein tunneling occurs without superradiance (the MSE converges). The
wavepacket calculations were complemented with exact numerical solutions
obtained by implementing a finite-difference code, leading to an excellent
agreement for rectangular and smooth barriers.

We have argued that while a stationary first quantized approach to the Klein
paradox has resulted in different and conflicting interpretations, a
time-dependent account adequately describes the dynamics. We
further believe such wavepacket calculations might be valuable in order to
have a qualitative or quantitative understanding for processes that should in
principle be described by spacetime QFT approaches, which are computationally
much more involved.

\vspace{.5cm}
%\begin{acknowledgments}
\textbf{Acknowledgments.} We would like to thank Rainer Grobe and Charles Su (Illinois State Univ.) for
helpful discussions. Financial support of MCIU, through the Grant
PGC2018-101355-B-100(MCIU/AEI/FEDER,UE) (XGdC, MP, DS), of Spanish MINECO,
project FIS2016-80681-P (MP), and of the Basque Government Grant no. IT986-16
(MP, DS) is gratefully acknowledged.
%\end{acknowledgments}

\appendix

\section{Relativistic wave equations}

\label{AppRWE}

The Klein-Gordon (KG) equation, describing spin-0 bosons of rest mass $m$, in
the presence of a time-independent potential (electrostatic type) in the
minimal coupling scheme is given in the canonical form by%
\begin{equation}
\hbar^{2}\partial_{t}^{2}\psi-c^{2}\hbar^{2}\partial_{x}^{2}\psi
+2iV(x)\hbar\partial_{t}\psi+\left(  m^{2}c^{4}-V(x)^{2}\right)  \psi=0
\label{cano}%
\end{equation}
We mostly use the Hamiltonian version of the KG\ equation, given by%
\begin{equation}
ih\partial_{t}\Psi=-\frac{\hbar^{2}}{2m}\left(  \sigma_{3}+i\sigma_{2}\right)
\partial_{x}^{2}\Psi+\left(  mc^{2}\sigma_{3}+V(x)\right)  \Psi\label{KG}%
\end{equation}
where $\Psi$ has components
\begin{equation}
\Psi=\left(
\begin{array}
[c]{c}%
\varphi\\
\chi
\end{array}
\right)  \label{comp}%
\end{equation}
linked to the solutions $\psi(t,x)$ of the canonical KG equation through%
\begin{align}
&  \psi(t,x)=\varphi(t,x)+\chi(t,x)\\
&  i\hbar\partial_{t}\psi-V(x)\psi=mc^{2}\left(  \varphi-\chi\right)  .
\end{align}
$\sigma_{j}$ with $j=1,2,3$ denotes the Pauli matrices. Recall \cite{greiner}
that the local charge probability density $\rho(t,x)$ can be negative; it is
given in the Hamiltonian formulation by%
\begin{equation}
\rho(t,x)=\varphi^{\ast}(t,x)\varphi(t,x)-\chi^{\ast}(t,x)\chi(t,x)
\label{KGcharge}%
\end{equation}
so that the positive and negative charge amplitudes are respectively carried
by $\varphi$ and $\chi$. The more general expression for the Klein-Gordon
scalar product for two wavefunctions $\Psi_{1}=\left(  \varphi_{1},\chi
_{1}\right)  $ and $\Psi_{2}=\left(  \varphi_{2},\chi_{2}\right)  $ given by
Eq.\ (\ref{comp}) is given by
\begin{equation}
\left\langle \Psi_{1}\right\vert \left.  \Psi_{2}\right\rangle _{KG}%
=\left\langle \Psi_{1}\right\vert \sigma_{3}\left\vert \Psi_{2}\right\rangle
=\int dx(\varphi_{1}^{\ast}\varphi_{2}-\chi_{1}^{\ast}\chi_{2}). \label{KGsp}%
\end{equation}

The Dirac equation for a fermion propagating in one spatial direction can be
reduced to an equation for a 2 component wavefunction $\Phi$. Hence in one dimension the spin
is ``frozen'' and does not play any role, but in more than one spatial dimensions, there could be spin flip upon reflection on the barrier.\ It can be seen either by starting from the usual Dirac equation
with an electrostatic potential,
\begin{equation}
[i \hbar(\gamma^{0} \partial_{t} + \vec{\gamma}.\vec{\nabla}) - \gamma^{0}
V(x) - m c^{2}] \Upsilon(t,x) =0
\end{equation}
where $\gamma^{\mu}$, $\mu=0,...,3$ are the usual $4 \times4$ gamma matrices
\cite{greiner}, and then restricting the 4 component wavefunction
$\Upsilon(t,x)$ to the non-trivial subspace involving only two components
(valid in the laboratory frame); alternatively one can start from a
non-covariant equation implementing the usual constraints leading to the Dirac
equation for one spatial dimension \cite{nitta}. This yields the 2-component
Dirac equation for $\Phi$
\begin{equation}
i\hbar\partial_{t}\Phi=-i\hbar c\sigma_{1}\partial_{x}\Phi+\left(
mc^{2}\sigma_{3}+V(x)\right)  \Phi. \label{D}%
\end{equation}
Recall that in the Dirac case the local probability density $\rho
(t,x)=\Phi^{\dag}(t,x)\Phi(t,x)$ is always positive and can be written as
$\rho(t,x)=\left\vert \varphi(t,x)\right\vert ^{2}+\left\vert \chi
(t,x)\right\vert ^{2}$ if we label the components as in Eq. (\ref{comp}). The
positive definite scalar product is
\begin{equation}
\left\langle \Phi_{1}\right\vert \left.  \Phi_{2}\right\rangle =\int
dx(\varphi_{1}^{\ast}\varphi_{2}+\chi_{1}^{\ast}\chi_{2}). \label{Dsp}%
\end{equation}

\section{Multiple Scattering Expansion amplitudes}

In the MSE one treats the barrier problem as multiple reflections involving
two steps. The barrier transmission and reflection amplitudes are expressed in
terms of the transmission and reflection amplitudes of each step with
different boundary conditions in order to obtain Eq (2) of the paper. We set
$\hbar=1$ in this Section.

\subsection{Klein-Gordon Equation}

For the rectangular barrier, the amplitudes are obtained from the left ($l$)
step $V\theta(x)$ and the right ($r$) step $V\theta(L-x)$. The amplitudes for
each step are obtained straightforwardly from the continuity of the wave
function in the canonical form and its first derivative at $x=0$ for the left
step and $x=L$ for the right step by considering a wave incoming from the left
($i$) or ``outcoming'' from the right ($o$). This gives the amplitudes%

\begin{equation}%
\begin{split}
t_{l}^{i} = \frac{2 p_{1}}{p_{1} + p_{2}(p_{1})} , \quad t_{r}^{i} = \frac{2
	p_{2}(p_{1})}{p_{1}+p_{2}(p_{1})} e^{i(p_{2}(p_{1})-p_{1})L},\\
r_{l}^{o} =\frac{p_{2}(p_{1})-p_{1}}{p_{2}(p_{1})+p_{1}}, \quad r_{r}^{i} =
r_{l}^{o}e^{2ip_{2}(p_{1})L}, \quad t_{l}^{o} = \frac{2p_{2}(p_{1})}%
{p_{1}+p_{2}(p_{1})}%
\end{split}
\end{equation}

For the smooth barrier given by  $V_{s}(x,\epsilon)=\frac{V}{2}\left[  \tanh(\epsilon x)-\tanh\left(
\epsilon(x-L)\right)  \right] $, the problem involves the two
smooth steps
\begin{equation}
\label{2steps}%
\begin{split}
V_{s1}(x)  &  = \frac{V}{2} \left(  1+ \tanh(\epsilon x)\right) \\
V_{s2}(x)  &  = \frac{V}{2} \left(  1+ \tanh(\epsilon(L-x))\right) \\
\end{split}
\end{equation}

The amplitudes for such hyperbolic tangent potentials can be extracted from
the ones given in Ref. \cite{rojas}. They are given by:
\begin{equation}%
\begin{split}
t_{l}^{i} &  =\frac{\Gamma(-i\nu+\lambda-i\mu)\Gamma(1-i\nu-\lambda-i\mu
	)}{\Gamma(1-2i\mu)\Gamma(-2i\nu)}\\
t_{l}^{o} &  =\frac{\Gamma(i\mu+\lambda+i\nu)\Gamma(1+i\mu-\lambda+i\nu
	)}{\Gamma(1+2i\nu)\Gamma(2i\mu)}\\
t_{r}^{i} &  =(t_{l}^{o})^{\ast}e^{2\epsilon i\mu L(\mu-\nu)}\\
r_{l}^{o} &  =\frac{\Gamma(-2i\mu)\Gamma(i\mu+\lambda+i\nu)\Gamma
	(1+i\mu-\lambda+i\nu)}{\Gamma(2i\mu)\Gamma(-i\mu+\lambda+i\nu)\Gamma
	(1-i\mu-\lambda+i\nu)}\\
r_{l}^{i} &  =\frac{\Gamma(2i\nu)\Gamma(-i\nu+\lambda-i\mu)\Gamma
	(1-i\nu-\lambda-i\mu)}{\Gamma(-2i\nu)\Gamma(i\nu+\lambda-i\mu)\Gamma
	(1+i\nu-\lambda-i\mu)}\\
r_{r}^{i} &  =(r_{l}^{o})^{\ast}e^{4i\epsilon\mu L}%
\end{split}
\end{equation}
with $\mu,\nu$ and $\lambda$ given by
\begin{gather}
\nu=\frac{p_{1}}{2\epsilon},\qquad\mu=\frac{p_{2}}{2\epsilon},\nonumber\\
\lambda=\frac{1}{2}+\frac{\sqrt{\epsilon^{2}-V_{0}^{2}}}{2\epsilon}%
\end{gather}

\subsection{Dirac Equation}

One can proceed similarly for the Dirac equation. For a rectangular barrier it
suffices to match the wavefunction at $x=0$ and $x=L$ (since the Dirac
equation is of first order in $x$) for the left step, and then the right step.
This yields, for positive energy waves (we drop the ``$+$'' superscript)
\begin{equation}%
\begin{split}
t_{l}^{i} = \frac{2 \alpha_{1}}{\alpha_{1} + \alpha_{2}} \frac{n_{1}}{n_{2}},
\quad t_{l}^{o} = \frac{2\alpha_{2}}{\alpha_{1}+\alpha_{2}}\frac{n_{2}}{n_{1}%
},\quad t_{r}^{i} = \frac{2 \alpha_{2}}{\alpha_{1}+\alpha_{2}} e^{i(p_{2}%
	(p_{1})-p_{1})L}\frac{n_{2}}{n_{1}}\\
r_{l}^{i} =\frac{\alpha_{1}-\alpha_{2}}{\alpha_{1}+\alpha_{2}}, \quad
r_{l}^{o} =\frac{\alpha_{2}-\alpha_{1}}{\alpha_{1}+\alpha_{2}}, \quad
r_{r}^{i} = r_{l}^{o} e^{2ip_{2}(p_{1})L}%
\end{split}
\end{equation}
Negative energy amplitudes can be obtained similarly (though we do not need
them in this work).

\section{Wavepackets}

Let the initial state be given by
\begin{equation}
G(0,x)=\left(
\begin{array}
[c]{c}%
\exp(\frac{-(x-x_{0})^{2}}{4d^{2}}-ip_{0}x/\hbar\\
0
\end{array}
\right)  \label{is1}%
\end{equation}
In the Klein-Gordon case, the wavepacket expansion at time $t$, given by Eq.
(7) of the paper, can be expanded as%
\begin{equation}%
\begin{split}
G_{KG}(t,x) &  =\theta_{1}\int dp_{1}\left(  c_{KG}^{+}(\mathcal{A}%
_{1}e^{ip_{1}x/\hbar}+\mathcal{B}_{1}^{{}}e^{-ip_{1}x/\hbar})e^{-i\sqrt
	{m^{2}c^{4}+p_{1}^{2}c^{2}}t/\hbar}+c_{KG}^{-}\mathcal{A}_{1}e^{ipx/\hbar
}e^{+i\sqrt{m^{2}c^{4}+p_{1}^{2}c^{2}}t/\hbar}\right)  \\
&  +\theta_{2}\int dp_{1}\left(  c_{KG}^{+}(\mathcal{A}_{2}^{{}}%
e^{ip_{2}(p_{1})x/\hbar}+\mathcal{B}_{2}^{{}}e^{-ip_{2}(p_{1})x/\hbar
})e^{-i\sqrt{m^{2}c^{4}+p_{1}^{2}c^{2}}t/\hbar}\right)  \\
&  +\theta_{3}\int dp_{1}\left(  c_{KG}^{+}\mathcal{A}_{3}^{{}}e^{ip_{3}%
	x/\hbar}e^{-i\sqrt{m^{2}c^{4}+p_{1}^{2}c^{2}}t/\hbar}\right)
\end{split}
\end{equation}
where  $\theta_{j}$ ensures each expression is used only in the region $j$ in
which it is valid ($\theta_{1}=\theta(-x),\theta_{2}=\theta(x)\theta(L-x)$ and
$\theta_{3}=\theta(x-L)$). The amplitudes $\mathcal{A}_{j}$ are defined by
\begin{equation}
\mathcal{A}_{j}^{\pm}=%
\begin{pmatrix}
mc^{2}\pm\sqrt{m^{2}c^{4}+p_{j}^{2}c^{2}}\\
mc^{2}\mp\sqrt{m^{2}c^{4}+p_{j}^{2}c^{2}}%
\end{pmatrix}
A_{j}^{\pm}%
\end{equation}
where $A_{j}^{\pm}$ are given by Eq. (2) of the paper. The amplitudes
$\mathcal{B}_{j}^{\pm}$ are related similarly to the amplitudes $B_{j}^{\pm}$
given by Eq. (2). Finally the expansion coefficients $c_{KG}^{\pm}=\left\langle \Psi_{1}^{\pm}\right\vert \left.
G\right\rangle _{KG}$ are
readily computed from the Klein-Gordon scalar product (\ref{KGsp}):%
\begin{equation}
c_{KG}^{\pm}=\frac{mc^{2}\pm\sqrt{m^{2}c^{4}+p_{1}^{2}c^{2}}}{\sqrt{m^{2}%
		c^{4}+p_{1}^{2}c^{2}}}\exp\left(  -d^{2}(p_{1}-p_{0})^{2}/\hbar^{2}-i\left(
p_{1}-p_{0}\right)  x_{0}/\hbar\right)
\end{equation}

For the Dirac case, we first recall the expression of the amplitudes and
spinor coefficients in the plane wave expression given by Eq.\ (4) of the
paper. The spinor coefficients are given, for each region $j\ $of the
rectangular barrier, by the following expressions:%
\begin{align}
\alpha_{j}^{+}(p_{1}) &  =\frac{p_{1}}{mc^{2}+\sqrt{m^{2}c^{4}+p_{1}^{2}c^{2}%
}}\text{ for }j=1,3\\
\alpha_{2}^{+}(p_{1}) &  =-\frac{\sqrt{p_{1}^{2}c^{2}-2V\sqrt{m^{2}c^{4}%
			+p_{1}^{2}c^{2}}+V^{2}}}{mc^{2}+\sqrt{m^{2}c^{4}+p_{1}^{2}c^{2}}-V}\text{ }\\
\alpha_{j}^{-}(p_{1}) &  =\frac{p_{1}}{mc^{2}-\sqrt{m^{2}c^{4}+p_{1}^{2}c^{2}%
}}\text{ for }j=1,3\\
\alpha_{2}^{-}(p_{1}) &  =-\frac{\sqrt{p_{1}^{2}c^{2}+2V\sqrt{m^{2}c^{4}%
			+p_{1}^{2}c^{2}}+V^{2}}}{mc^{2}-\sqrt{m^{2}c^{4}+p_{1}^{2}c^{2}}-V}.\text{ }%
\end{align}
The amplitudes $A_{j}$ and $B_{j}$ are then obtained by using the connection
formulas at $x=0$ and $x=L$ (since the MSE converges), yielding
\begin{gather}
B_{1}=\frac{(\alpha_{1}-\alpha_{2})(\alpha_{1}+\alpha_{2})\sin(Lp_{2}%
	(p))}{2i\alpha_{1}\alpha_{2}\cos(Lp_{2}(p_{1}))+({\alpha_{1}}^{2}+{\alpha_{2}%
	}^{2})\sin(Lp_{2}(p))}\\
A_{2}=\frac{2\alpha_{1}(\alpha_{1}+\alpha_{2})}{(\alpha_{1}+\alpha_{2}%
	)^{2}-(\alpha_{1}-\alpha_{2})^{2}e^{2iLp_{2}(p_{1})}}\frac{n_{1}}{n_{2}}\\
B_{2}=\frac{2\alpha_{1}(\alpha_{1}-\alpha_{2})e^{2iLp_{2}(p_{1})}}%
{-(\alpha_{1}+\alpha_{2})^{2}+(\alpha_{1}-\alpha_{2})^{2}e^{2iLp_{2}(p_{1})}%
}\frac{n_{1}}{n_{2}}\\
A_{3}=\frac{2i\alpha_{1}\alpha_{2}e^{-iLp_{1}}}{2i\alpha_{1}\alpha_{2}%
	\cos(Lp_{2})+({\alpha_{1}}+{\alpha_{2}})^{2}\sin(Lp_{2})},\label{coedir}%
\end{gather}
where $n_{j}^{{}}$ are normalization coefficients. Finally, when the initial
state is given by Eq. (\ref{is1}), we need the expansion coefficients
$c_{D}^{\pm}=\left\langle \Phi_{1}^{\pm}\right\vert
\left.  G\right\rangle $ in order to compute the time-dependent wavepacket $G(t,x)$ given
by Eq. (9) of the paper. These coefficients are obtained trough the Dirac
scalar product (\ref{Dsp}) as%
\begin{equation}
c_{D}^{\pm}=n^{\pm}(p_{1})\exp\left(  -d^{2}(p_{1}-p_{0})^{2}/\hbar
^{2}-i\left(  p_{1}-p_{0}\right)  x_{0}/\hbar\right)
\end{equation}
with $n^{\pm}(p_{1})=\left(  1+\alpha_{1}^{\pm}(p_{1})^{2}\right)  ^{-1/2}.$

\section{Finite-Difference Computations}

For the Klein-Gordon equation, real-space approaches are now possible
\cite{keitel,bauke} although they are computationally more demanding than the
Fourier-transformed based split operator methods. The evolution operator for
an infinitesimal time step $\delta t$ is given by
\begin{equation}
U_{KG}(\delta t)=e^{-i\hat{H}_{KG}\delta t/\hbar}%
\end{equation}
where
\begin{equation}
\hat{H}_{KG}=-\frac{\hbar^{2}}{2m}\left(  \sigma_{3}+i\sigma_{2}\right)
\partial_{x}^{2}+\left(  mc^{2}\sigma_{3}+V(x)\right)  \label{hamkg}%
\end{equation}
is obtained from the right hand-side of Eq. (\ref{KG}). This operator is
pseudo-unitary in the sense that $\sigma_{3}U_{KG}^{\dagger}\sigma_{3}%
U_{KG}=\mathbb{I}.$ It is given explicitly by
\begin{equation}
U_{KG}(\delta t)=e^{-i\hat{H}_{KG}\delta t}=%
\begin{pmatrix}
\exp(-imc^{2}\delta t) & 0\\
0 & \exp(imc^{2}\delta t)
\end{pmatrix}
\sum_{n=0}^{n_{\mathrm{max}}}{\frac{1}{n!}u_{KG}^{n}} \label{Ukg}%
\end{equation}
where $u_{KG}$ is given by Eq. (\ref{ukg}) and $n_{\mathrm{max}}$ sets the
number of terms kept in the Taylor expansion.

To solve the Klein-Gordon equation, Eq. (\ref{KG}), numerically, we use the
finite-difference fourth order approximation of the second derivative with
respect to position.
%, $\partial^2_x$ in the Hamiltonian, Eq. (\ref{hamkg}).
This approximation is given by:
\begin{equation}
\partial_{x}^{2}=\frac{-f(x-2\delta x)+16f(x-\delta x)-30f(x)+16f(x+\delta
	x)-f(x+2\delta x)}{12\delta x^{2}}+O(\delta x^{4}). \label{fdapp}%
\end{equation}
The one dimensional space of length $l$ is discretized into a lattice of
$N_{x}$ points, $X=\{-\frac{l}{2}+n\delta x\ ;n=0,1,2,\dots Nx\}$ with $\delta
x=l/N_{x}$. Thus, this second derivative is represented by an $N_{x}\times
N_{x}$ matrix:
\begin{equation}
\partial_{x}^{2}=%
\begin{pmatrix}
& -30 & 16 & -1 & 0 & 0 & \dots\\
& 16 & -30 & 16 & -1 & 0 & \dots\\
& -1 & 16 & -30 & 16 & -1 & \ddots\\
& 0 & -1 & 16 & -30 & 16 & \ddots\\
& 0 & 0 & -1 & 16 & -30 & \ddots\\
& 0 & 0 & 0 & -1 & 16 & \ddots\\
& \vdots & \vdots & \vdots & \ddots & \ddots & \ddots\\
&  &  &  &  &  &
\end{pmatrix}
. \label{secorderderiva}%
\end{equation}
The potential $V(x)$ is represented by a diagonal matrix $N_{x}\times N_{x}$
where the diagonal $V(X)$ is obtained by evaluating the potential $V(x)$ at
every point of the lattice $X$. We then use Eq. (\ref{secorderderiva}) and
$V(X)$ to construct the operator
\begin{equation}
u_{KG}=\delta t%
\begin{pmatrix}
i\frac{\hbar}{2m}\partial_{x}^{2}-iV(x) & i\frac{\hbar}{2m}\partial^{2}\\
-i\frac{\hbar}{2m}\partial_{x}^{2} & -i\frac{\hbar}{2m}\partial_{x}^{2}-iV(x)
\end{pmatrix}
. \label{ukg}%
\end{equation}

Using $u_{KG}$ and the matrix:
\begin{equation}
e^{- i \sigma_{3} m c^{2} \delta t /\hbar} =
\begin{pmatrix}
exp(-i m c^{2} \delta t) & 0\\
0 & exp(i m c^{2} \delta t)\\
&
\end{pmatrix}
\end{equation}
that commutes with both $\partial^{2}_{x}$ and $V(X)$, we obtain the time
evolution operator $U_{KG}(\delta t) = e^{- i \hat{H}_{KG} \delta t}$ given by
Eq. (\ref{Ukg}). This operator is represented by a $2N_{x} \times2N_{x}$
matrix that is applied recursively on the one-dimensional vector $\Psi(t,X)$
to obtain $\Psi(t+dt,X)$ following
\begin{equation}
\Psi(t+ \delta t,x) =%
\begin{pmatrix}
\varphi(t+\delta t,x)\\
\chi(t+\delta t,x)
\end{pmatrix}
= U(\delta t)
\begin{pmatrix}
\varphi(t,x)\\
\chi(t,x)
\end{pmatrix}
\end{equation}
The numerical stability criteria was implemented similarly as in Ref.
\cite{bauke}.

The numerical solutions to the Dirac equation were obtained along the same
procedure. An initial Dirac wavepacket $G(t=0,x)$ is evolved numerically
according to Eq. (\ref{D}) by implementing the finite-difference approximation
of the Hamiltonian
\begin{equation}
\hat{H}_{D}=-i\hbar c\sigma_{1}\partial_{x}+\left(  mc^{2}\sigma
_{3}+V(x)\right)  \label{hamd}%
\end{equation}
The time evolution operator is then obtained as
\begin{equation}
U_{D}(\delta t)=e^{-i\hat{H}_{D}\delta t}=%
\begin{pmatrix}
\exp(-imc^{2}\delta t) & 0\\
0 & \exp(imc^{2}\delta t)
\end{pmatrix}
\sum_{n=0}^{n_{\mathrm{max}}}{\frac{1}{n!}u_{D}^{n}}, \label{UD}%
\end{equation}
where $u_{D}$ is given by Eq. (\ref{uD}). The time evolution operator, Eq.
(\ref{UD}) was built using the fifth-order finite-difference approximation of
the first spatial derivative
\begin{equation}
\partial_{x}f(x)\approx\frac{f(x-2\delta x)-8f(x-\delta x)+8f(x+\delta
	x)-f(x)}{12\delta x}+O(\delta x^{4})
\end{equation}
This approximation is employed on a discretized lattice of dimension $l$,
yielding the matrix:
\begin{equation}
\partial_{x}=%
\begin{pmatrix}
& 0 & 8 & -1 & 0 & 0 & \dots\\
& -8 & 0 & 8 & -1 & 0 & \dots\\
& 1 & -8 & 0 & 8 & -1 & \ddots\\
& 0 & 1 & -8 & 0 & 8 & \ddots\\
& 0 & 0 & 1 & -8 & 0 & \ddots\\
& 0 & 0 & 0 & 1 & -8 & \ddots\\
& \vdots & \vdots & \ddots & \ddots & \ddots & \ddots\\
&  &  &  &  &  &
\end{pmatrix}
\label{secorderderiva}%
\end{equation}
Along with the potential $V(X)$ (that is again diagonal) we obtain the matrix
representing the operator $u_{D}$,
\begin{equation}
u_{D}=-\delta t%
\begin{pmatrix}
iV(x) & \partial_{x}\\
\partial_{x} & iV(x)\\
&
\end{pmatrix}
\label{uD}%
\end{equation}
and finally the Dirac time evolution operator as per Eq. (\ref{UD}). Note that
the `fermion-doubling\textquotedblright\ problem \cite{muller} (appearance of
spurious numerical solutions to the Dirac equation) is avoided by working with
small space steps so that the momentum cutoff remains well below the largest
momentum included in the wavepackets.

In the computations displayed in Fig. 1 of the paper, we used a lattice of
length $l=4000$ and $N_{x}=3\times10^{6}$ points. The time step was $\delta t
= 10^{-3}$. For the finite-difference solutions to the Dirac equation
displayed in Fig. 2 of the paper we used $l=4000$, $N_{x}=5\times10^{5}$ and
$\delta t = 2\times10^{-2}$. The momentum cutoff, $\pi/(l/N_{x})$, is 2 orders
of magnitude larger than the highest momentum contributing to the wavepacket.


\begin{thebibliography}{99}
\bibitem {units}We use natural units $\hbar=m=c=\varepsilon_{0}=1$ where
$\varepsilon_{0}$ is the vacuum permittivity. For the Dirac equation, $m$ is
the electron mass. For the Klein-Gordon equation, the conversion to SI units
depends on the specific boson mass $m$, eg for a pion meson $\pi^{+}$ the mass
entering the conversion factor is 139.57 MeV/c$^{2}$.
\end{thebibliography}
\end{document}